\documentclass[final,unsortedaddress,twocolumn,floatfix,%
  prb,amsmath,amssymb,showpacs]{revtex4}

\usepackage{amsmath}
\usepackage{amssymb}
\usepackage{braket}
\usepackage{color}
\usepackage{epsfig}
\usepackage{graphics}
\usepackage{latexsym}
\usepackage{multirow}
\usepackage{natbib}

\begin{document}

\title{
  Non--Adiabatic Contributions to the Free Energy from the Electron--Phonon
  Interaction for Na, K, Al, and Pb
}

\date{\today}

\author{N.~Bock}
\email{nbock@lanl.gov}

\affiliation{
  Theoretical Division, Los Alamos National Laboratory, Los Alamos, New Mexico
  87545
}

\author{D.~Coffey}

\affiliation{
  Dept. of Physics, Buffalo State College, Buffalo, New York 14222
}

\author{Duane~C.~Wallace}

\affiliation{
  Theoretical Division, Los Alamos National Laboratory, Los Alamos, New Mexico
  87545
}


\begin{abstract}

We calculate the non--adiabatic contributions to the free energy of metals due
to the electron--phonon interaction at intermediate temperatures, $0 \leqslant
k_{B} T < \epsilon_{F}$ for four different nearly free electron metals, Na, K,
Al, and Pb. We calculate its value for $T = 0$ which has not been calculated
before and we study its low--temperature behavior.

\vspace{0.2cm}

\noindent
LA-UR-04-8709

\end{abstract}

\maketitle

\section{
  Introduction
}

For many practical applications one needs accurate values of thermodynamic
properties of solids and liquids both at low pressures, where direct
measurements are available, and in regions of temperature and pressure where
data are absent. This requires a model Hamiltonian for the system which gives
accurate results for the Helmholtz free energy. For a metal crystal the free
energy consists of three contributions,

\begin{eqnarray}
  F & = & \Phi_{0} (V) + F_{E} (V, T) + F_{I} (V, T)
    \nonumber \\
  & = & \Phi_{0} (V) + F_{el} (V, T) + F_{ep} (V, T)
    \nonumber \\
  \label{eq_free_energy}
  & & \mbox{} + F_{ph} (V, T) + F_{anh} (V, T).
\end{eqnarray}

\noindent
$\Phi_{0} (V)$, the static lattice potential, represents the total energy when
the ions are located at lattice sites and the electrons are in their ground
state. $F_{I} (V, T)$ is the free energy from ion vibrations, and $F_{E} (V, T)$
is the free energy associated with thermal excitation of electrons from their
ground state. $F_{I} (V, T)$ is the dominant temperature contribution. It
consists of the quasi--harmonic phonon contribution, $F_{ph} (V, T)$, plus the
small anharmonic term $F_{anh} (V, T)$ which expresses phonon--phonon
interactions. $F_{E} (V, T)$ consists of $F_{el} (V, T)$ representing the
thermal excitation of independent electrons, plus $F_{ep} (V, T)$, the
contribution from interactions between electronic excitations and phonons.

In this paper we determine the magnitude and temperature dependence of the
non--adiabatic contributions to the free energy from the electron--phonon
interaction for different metals and compare with previous treatments.

Our calculations are done for a constant density $\rho$ to eliminate concern for
the density dependence of phonon frequencies and electron--phonon interaction
matrix elements. The density is that at the temperature $T_{\rho}$, where the
phonon frequencies are measured. The melting temperature at this density is
higher than the customary zero--pressure melting temperature. Our calculations
cover the range from $T = 0$ to above $T_{m}$.

The paper is organized as follows: In section \ref{sec_ep_free_energy} we will
introduce the expressions used for the electron--phonon part of the free energy.
In section \ref{sec_results} our results are presented with a more detailed
discussion of the numerical methods used.

\section{
  Electron--Phonon Free Energy
}
\label{sec_ep_free_energy}

\subsection{
  Analytic Form of the Free Energy
}

Let us briefly discuss the physical origin of the contributions to $F_{ep}$, and
show in the process that our formulation is free of double--counting errors. In
electronic structure theory we calculate the electronic ground state energy with
static nuclei, as a function of the nuclear positions, $\left\{ \vec{r}_{K}
\right\} = \vec{r}_{K = 1, \cdots, N}$. This energy is the ground state
adiabatic potential $\Phi \left( \left\{ \vec{r}_{K} \right\} \right)$, which is
resolved into $\Phi_{0} + \Phi_{ph} + \Phi_{anh}$, and which together with the
nuclear kinetic energy gives the free energy contribution $\Phi_{0} + F_{ph} +
F_{anh}$ in eq.  (\ref{eq_free_energy}). We next do the same calculation for the
excited electronic states, labeled $n = 1, 2, \cdots$ of the energy $E_{n}
\left( \left\{ \vec{r}_{K} \right\} \right)$ with static nuclei. Here only the
\emph{excitation} energies $E_{n} - \Phi$ are considered since $\Phi$ is already
included in the nuclear motion contributions. Excited electronic states with
nuclei at the crystal lattice sites are calculated in electronic structure
theory, and their energy levels are expressed in the electronic density of
states, which provides the free energy contribution $F_{el}$ in eq.
(\ref{eq_free_energy}). Next the thermally averaged vibrational contributions to
the excited energies $E_{n} - \Phi$ are calculated, and these contributions
yield the adiabatic part $F^{ad}$ of the electron--phonon free energy, eq.
(\ref{eq_F_ad}). Finally the non--adiabatic correction to all the electronic
energy levels, ground and excited states alike, is calculated by allowing the
nuclear motion to mix the electronic states. This produces the terms $F_{1}^{na}
+ F_{2}^{na}$, eqs. (\ref{eq_F_1_na}) and (\ref{eq_F_2_na}). The non--adiabatic
part can formally be attributed to the terms in the trace of the partition
function in which the nuclear kinetic energy operator operates on the electronic
wavefunctions. A detailed discussion of the resolution of the crystal
Hamiltonian, and the corresponding free energy contributions may be found in
\citet{Wallace_02:SPCL} Secs. 4 and 18 and pp.  91 -- 94.

While this theory is valid for metals in general, an approximation is available
from which we can calculate $F_{ep}$ from previously calibrated models for the
nearly--free electron metals. This is pseudopotential perturbation theory, which
starts from free electrons in zeroth order, and treats the screened
electron--ion interaction (the pseudopotential) as a perturbation. Electronic
bandstructure effects then arise in standard perturbation theory. The
electron--phonon theory becomes a double perturbation expansion, in the
pseudopotential and in the displacement of nuclei from equilibrium. But the
displacement expansion converts to a pseudopotential expansion, and the leading
contributions to $F_{ep}$ are all of second order in the pseudopotential. The
derivation may be found in \cite[Sec. 18]{Wallace_02:SPCL}.

The electron--phonon contribution is written as a sum over the three terms
mentioned in the previous paragraph. They are

\begin{equation}
  \label{eq_F_ep}
  F_{ep} = F^{ad} + F_{1}^{na} + F_{2}^{na},
\end{equation}

\noindent
where the single contributions are given by

\begin{widetext}
\begin{eqnarray}
  \label{eq_F_ad}
  \frac{F^{ad}}{N} & = & \sum_{\vec{p} \vec{k} \vec{Q} \lambda}
  \frac{\hbar^{2}}{N^{2} M}
  \frac{n_{\vec{k} \lambda} + \frac{1}{2}}{\hbar \omega_{\vec{k} \lambda}}
  \left( f_{\vec{p}} - g_{\vec{p}} \right)
  \left\{
    \frac{\left[ \left( \vec{k} + \vec{Q} \right) \cdot
      \hat{\eta}_{\vec{k} \lambda} \right]^{2}
      \left[ U \left( \vec{k} + \vec{Q} \right) \right]^{2} }
      {\epsilon_{\vec{p}} - \epsilon_{\vec{p} + \vec{k} + \vec{Q}}}
    -
    \frac{\left[ \vec{Q} \cdot \hat{\eta}_{\vec{k} \lambda} \right]^{2}
      \left[ U \left( \vec{Q} \right) \right]^{2} }
      {\epsilon_{\vec{p}} - \epsilon_{\vec{p} + \vec{Q}}}
  \right\} \\
  \label{eq_F_1_na}
  \frac{F_{1}^{na}}{N} & = & \sum_{\vec{p} \vec{k} \vec{Q} \lambda}
  \frac{\hbar^{2}}{N^{2} M}
  \hbar \omega_{\vec{k} \lambda}
  \left( n_{\vec{k} \lambda} + \frac{1}{2} \right)
  \frac{f_{\vec{p}}}{\epsilon_{\vec{p}} - \epsilon_{\vec{p} + \vec{k} + \vec{Q}}}
  \frac{\left[ \left( \vec{k} + \vec{Q} \right) \cdot
    \hat{\eta}_{\vec{k} \lambda} \right]^{2}
    \left[ U \left( \vec{k} + \vec{Q} \right) \right]^{2}}
    {\left[ \epsilon_{\vec{p}} - \epsilon_{\vec{p} + \vec{k} + \vec{Q}} \right]^{2}
     - \left[ \hbar \omega_{\vec{k} \lambda} \right]^{2}} \\
  \label{eq_F_2_na}
  \frac{F_{2}^{na}}{N} & = & \sum_{\vec{p} \vec{k} \vec{Q} \lambda}
  \frac{\hbar^{2}}{2 N^{2} M}
  f_{\vec{p}} \left( 1 - f_{\vec{p} + \vec{k} + \vec{Q}} \right)
  \frac{\left[ \left( \vec{k} + \vec{Q} \right) \cdot
    \hat{\eta}_{\vec{k} \lambda} \right]^{2}
    \left[ U \left( \vec{k} + \vec{Q} \right) \right]^{2}}
    {\left[ \epsilon_{\vec{p}} - \epsilon_{\vec{p} + \vec{k} + \vec{Q}} \right]^{2}
     - \left[ \hbar \omega_{\vec{k} \lambda} \right]^{2}}.
\end{eqnarray}
\end{widetext}

\noindent
Here and in the following we will calculate and quote our results as per atom.
$f_{\vec{p}}$ is the Fermi--Dirac distribution function at finite temperature
and $g_{\vec{p}}$ is the same at $T = 0$. $n_{\vec{k} \lambda}$ is the
Bose--Einstein distribution function at finite temperature and
$\hat{\eta}_{\vec{k} \lambda}$ is the polarization vector of the phonon branch
$\lambda$ for wave vector $\vec{k}$ which is inside the Brillouin zone.
$\vec{Q}$ is a reciprocal lattice vector and $\omega_{\vec{k} \lambda}$ is the
frequency of a phonon mode. $U (\vec{k} + \vec{Q})$ is the Fourier transform of
the pseudopotential for momentum transfer $\vec{k} + \vec{Q}$. In our
calculations we use two models for the pseudopotential, the Harrison and the
Ashcroft models, which are screened by exchange and electron--electron
interactions (detailed forms are given in \citet[pp. 312]{Wallace_72:TOC}). The
pseudopotential parameters are listed in Table
\ref{table_pseudopotential_parameters}. $\epsilon_{\vec{p}}$ is the electron
energy measured relative to the Fermi energy.

In the above formulas, every term is of second order in $U (q)$. The appearance
of free electron energies in the denominators results from the use of
perturbation theory. The pseudopotential modification of the electron
wavefunctions gives rise to the last term in braces in eq. (\ref{eq_F_ad}). An
alternate derivation, by means of Matsubara frequencies, is outlined in the
appendix.

This is the form of the electron--phonon contribution to the free energy
originally calculated by \citet{Eliashberg_60_JETP} who studied the
electron--phonon interaction in the superconducting state. It is known that the
non--adiabatic contribution dominates the adiabatic one at low temperature
\cite{Allen_78} and we will focus on $F_{1,2}^{na}$ in this paper. $F^{ad}$ will
be examined in a future publication.

\subsection{
  \label{sec_analytic_temp_dependence}
  Analytic temperature dependences
}

From eqs. (\ref{eq_F_1_na}) and (\ref{eq_F_2_na}) we estimate the low--$T$
dependence of the two non--adiabatic contributions with the leading order terms
of a Sommerfeld expansion \cite[pp. 45]{Ashcroft_76:SSP}. This is certainly
accurate for nearly free electron metals which have smooth density of states
around the Fermi surface, but is less certain for transition metals and
actinides because their electronic density of states tends to fluctuate strongly
around the Fermi surface which makes power series expansions around
$\epsilon_{F}$ less accurate.

\subsubsection{
  $F_{1}^{na}$
}

In order to analyze the first non--adiabatic contribution we rewrite the
Fermi--Dirac factor as follows:

\begin{equation}
  \label{eq_f}
  f_{\vec{p}} =  g_{\vec{p}} + \left( f_{\vec{p}} - g_{\vec{p}} \right).
\end{equation}

\noindent
At high temperatures we can expand the other temperature dependent factor

\begin{equation}
  \label{eq_n_plus_half}
  \hbar \omega \left( n_{\vec{k}} + \frac{1}{2} \right)
  =
  \hbar \omega
  \left\{ \frac{1}{\beta \hbar \omega}
    + \frac{\beta \hbar \omega}{12}
    + \frac{\left( \beta \hbar \omega \right)^{3}}{720}
    + \cdots
  \right\}
\end{equation}

\noindent
in powers of $T$ and get a linear $T$ dependence in leading order. At low
temperature, the factor eq. (\ref{eq_n_plus_half}) will go over to a constant
and we find that the temperature dependence of the ground state part of
$F_{1}^{na}$ given by the first term in eq. (\ref{eq_f}) will be given by

\begin{equation}
  \label{eq_T_dependence}
  F_{1}^{na} = \left\{
    \begin{tabular}{l}
      const. at $T = 0$ \\
      $k T$ at high $T$
    \end{tabular}
    \right.
\end{equation}

\noindent
The contribution from the term in parentheses in eq. (\ref{eq_f}) is restricted
in phase space to a small volume around the Fermi surface, whereas the ground
state contribution is restricted only to $p < p_{F}$. Since the $\left(
f_{\vec{p}} - g_{\vec{p}} \right)$ contribution has much less phase space than
the $g_{\vec{p}}$ contribution, the ground state contribution will dominate at
all temperatures and we will not see any additional temperature dependence
beyond eq. (\ref{eq_T_dependence}).

Considering the factor $\hbar \omega / \left( \epsilon_{\vec{p}} -
\epsilon_{\vec{p'}} \right)$ in $F_{1}^{na}$ we expect this contribution to be
of order $\hbar \Braket{\omega} / \epsilon_{F}$ smaller than $F_{2}^{na}$ and we
will be able to neglect it. We calculated $F_{1}^{na}$ for Na to verify these
estimates and present it in section \ref{sec_results}.

\subsubsection{
  $F_{2}^{na}$
}

Taking a closer look at the integrand of the second non--adiabatic contribution
we notice that all factors but one are positive for all points in the
integration region. Only the energy difference in the denominator can have
different signs depending on the momenta $\vec{p}$ and $\vec{p'} = \vec{p} +
\vec{k} + \vec{Q}$.  We estimate its sign by dividing the integration phase
space into two regions

\begin{equation}
  \left| \Delta \epsilon \right| = \left| \epsilon_{\vec{p}} - \epsilon_{\vec{p'}} \right|
  \left\{
    \begin{array}{l}
      < \hbar \omega \mbox{\hspace{0.5cm}(Region 1)} \\
      > \hbar \omega \mbox{\hspace{0.5cm}(Region 2)}
    \end{array}
  \right.
\end{equation}

\noindent
At zero temperature due to the Fermi--Dirac factor, $f_{\vec{p}} \left( 1 -
f_{\vec{p'}} \right)$, the width of the phase space for Region 1 is twice the
phonon energy, $2 \hbar \omega$, in energy around the Fermi surface and very
small compared to Region 2 which extends everywhere outside Region 1.
$F_{2}^{na} (T = 0)$ will therefore be positive. Increasing temperature will
widen the phase space for Region 1 because of the softening of the Fermi surface
and will increase its negative contribution to the total but will leave the
contribution from Region 2 mostly unaffected. The overall effect will be to
reduce the  magnitude  of $F_{2}^{na}$, the more so with increasing temperature
but we expect $F_{2}^{na}$ to remain positive.

\begin{figure}
  \psfig{file=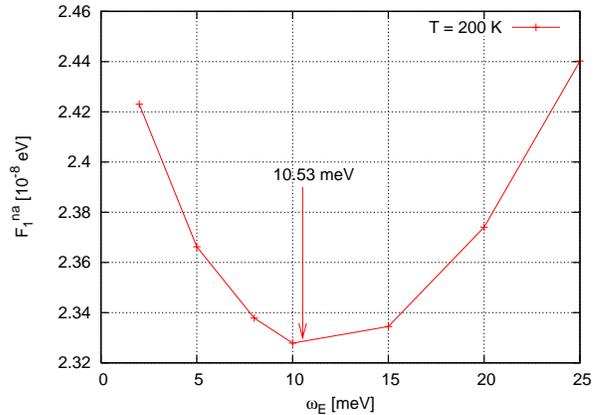,angle=-90,width=8cm}
  \caption{
    \label{fig_set_03}
    $\omega_{E}$--dependence of $F_{1}^{na}$ for Na at $T$ = 200 K.
  }
\end{figure}

From a Sommerfeld expansion of the Fermi--Dirac factor for small temperatures we
estimate the low--temperature dependence of $F_{2}^{na}$ to be given by

\begin{equation}
  \label{eq_2_na_T_dependence}
  F_{2}^{na} = C_{2} + A_{2} T^{2} + \cdots
\end{equation}

\noindent
Previous work on the electron--phonon contribution to the free energy focused on
calculating the specific heat and missed the factor $C_{2}$ which we will
calculate in section \ref{sec_results}. An extensive survey of the theoretical
calculations is given by Grimvall \cite[Tables III--VI]{Grimvall_76}. We will
also calculate the factor $A_{2}$ and compare with experiment and previous
estimates for different materials.

\begin{figure}
  \psfig{file=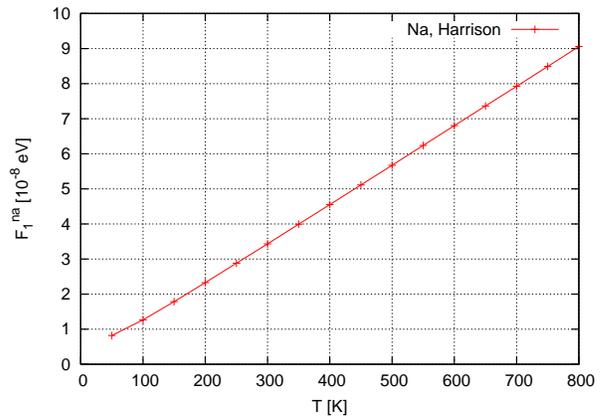,angle=-90,width=8cm}
  \caption{
    \label{fig_set_06}
    $F_{1}^{na}$ for Na.
  }
\end{figure}

As has been shown by previous authors, the low temperature specific heat of
normal metals has a $T^{3} \ln T$ contribution which comes from
electron--electron interaction, both those induced by phonons and those due to
the Coulomb interaction. Coffey and Pethick \cite{Coffey_Pethick_88} and Danino
and Overhauser \cite{Danino_Overhauser_82} for instance derived this
contribution for a Debye model. We are using an Einstein model in our
calculations of the non--adiabatic parts and can not pick up the $T^{3} \ln T$
term since it depends on the existence of acoustic phonon modes. It is known
however that this contribution is small compared to the others and we will not
calculate it.

\section{
  Results
}
\label{sec_results}
  
\subsection{
  Einstein Approximation
}

In this section we want to study the two non--adiabatic contributions using an
Einstein model. We will try to answer two main questions: Will we get the
correct temperature dependence from such a simplified model? How accurate will
our results be?

The details of the phonon model enter the single contributions through the
factor

\begin{equation}
  \left[ \left( \vec{k} + \vec{Q} \right)
  \cdot \hat{\eta}_{\vec{k} \lambda} \right]^{2}
\end{equation}

\noindent
in the integrands and through their explicit dependence on the phonon
frequencies, $\omega_{\vec{k} \lambda}$. The former contains details of the
phonon spectrum in terms of the polarization vectors of the phonon branches and
turns out to have a fairly weak effect on the integral, whereas the effect of
the latter can be very strong for acoustic phonon branches. When $\vec{k}$ goes
to zero at the zone center \footnote{We remind the reader that $\vec{k}$ is
restricted to the Brillouin zone.} the phonon frequencies of acoustic branches
vanish. If the integrand plus volume factors of $k^{2}$ is well--behaved and
finite for $\omega_{\vec{k} \lambda} \rightarrow 0$ we can expect to be able to
approximate the integral using a suitably chosen Einstein mode. If the integrand
plus volume factors diverges for vanishing $\omega_{\vec{k} \lambda}$ the
details of the phonon spectrum will be important and we can not expect to be
able to approximate the full phonon spectrum with a single frequency. There
might be a situation of course in which a single phonon frequency will result in
the correct temperature dependence, but without calculating the integral using a
realistic phonon dispersion we have no way of knowing what that single frequency
should be.

\begin{figure}
  \psfig{file=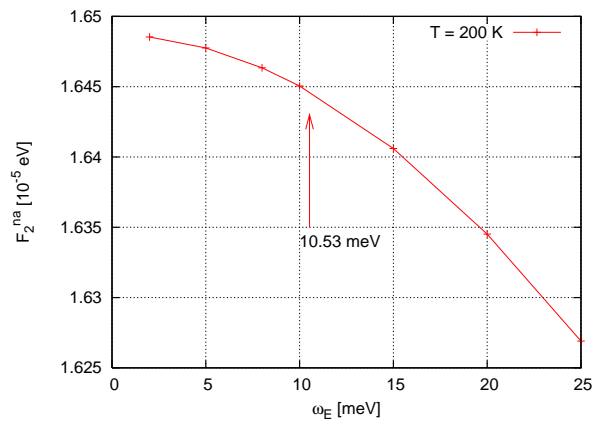,angle=-90,width=8cm}
  \caption{
    \label{fig_set_04}
    $\omega_{E}$--dependence of $F_{2}^{na}$ for Na at $T$ = 200 K.
  }
\end{figure}

Both non--adiabatic contributions, $F_{1,2}^{na}$, have the factor

\begin{equation}
  \frac{1}
  {
    \left[ \epsilon_{\vec{p}}
    - \epsilon_{\vec{p} + \vec{k} + \vec{Q}} \right]^{2}
    - \left[ \hbar \omega_{\vec{k} \lambda} \right]^{2}
  },
\end{equation}

\noindent
in common. At the zone center this turns into

\begin{equation}
  \frac{1}
  {
    \left[ \epsilon_{\vec{p}}
    - \epsilon_{\vec{p} + \vec{k} + \vec{Q}} \right]^{2}
  }
\end{equation}

\noindent
which is well behaved \footnote{We are working with a parabolic band and at
$\Gamma$--points with $\vec{Q} \neq 0$ the electronic energies
$\epsilon_{\vec{p}} \neq \epsilon_{\vec{p} + \vec{Q}}$.}. The additional factor
  
\begin{equation}
  \lim_{\omega_{\vec{k} \lambda} \rightarrow 0}
  \hbar \omega_{\vec{k} \lambda}
  \left( n_{\vec{k} \lambda} + \frac{1}{2} \right)
  =
  k_{B} T
\end{equation}

\noindent
in $F_{1}^{na}$ is also well behaved at the $\Gamma$ point. We conclude that a
single phonon frequency will reproduce $F_{1,2}^{na}$ with the correct
temperature dependence. We also expect it to be accurate and we will quantify
this last statement later in this paper.

Figs. \ref{fig_set_03} and \ref{fig_set_04} show the non--adiabatic
contributions for an Einstein model at $T$ = 200 K. Our a priori estimate for
the Einstein frequency is $\omega_{E} = \Braket{ \omega }$, where $\Braket{
\omega }$ is the average over the Brillouin zone. For reference we included our
estimates for the average phonon frequency in the graphs and in Table
\ref{table_material_constants}. As expected, the frequency dependence is very
weak and we can neglect it.

\begin{figure}
  \psfig{file=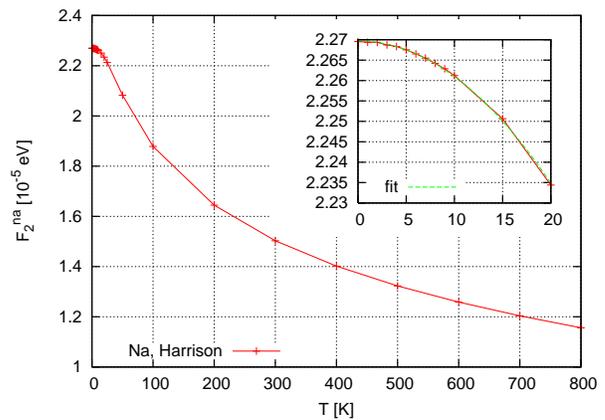,angle=-90,width=8cm}
  \caption{
    \label{fig_set_05}
    $F_{2}^{na}$ for Na. The inset shows the low temperature region.
  }
\end{figure}

\subsection{
  Numerical Techniques
}
\label{sec_numerical_techniques}

The two sums over the electron and phonon momenta, $\sum_{\vec{p}}$ and
$\sum_{\vec{k} \vec{Q}}$, are over all of reciprocal space. Assuming a large
sample size we therefore have the option to convert one or both of these sums
into integrals, using the well known relationship

\begin{equation}
  \sum_{\vec{p}}
  =
  \frac{ N V_{A} }{ \left( 2 \pi \right)^{3} }
  \int d^{3} p,
\end{equation}

\noindent
where $N$ is the number of ions in the sample, and $V_{A}$ the volume per ion.
In general, due to the completeness of the phonon eigenvectors in 3 dimensional
space,

\begin{equation}
  \label{eq_mode_sum}
  \sum_{\lambda} \left[ \left( \vec{k} + \vec{Q} \right)
    \cdot \hat{\eta}_{\vec{k} \lambda} \right]^{2}
  =
  \left| \vec{k} + \vec{Q} \right|^{2}.
\end{equation}

\noindent
In an Einstein model the sum over phonon branches can be performed outside the
integral since the only factors that depend on the branch index, $\lambda$, are
contained in eq. (\ref{eq_mode_sum}). This means that the integrand does not
depend on the crystal structure. If we combine the double sum $\sum_{\vec{k}
\vec{Q}}$ with $\sum_{\vec{p'}}$ where $\vec{p'} = \vec{p} + \vec{k} + \vec{Q}$
we can replace both sums, $\sum_{\vec{p} \vec{p'}}$, in the two non--adiabatic
contributions with integrals without loss of generality. Eqs. (\ref{eq_F_1_na})
and (\ref{eq_F_2_na}) can then be written as

\begin{figure}
  \psfig{file=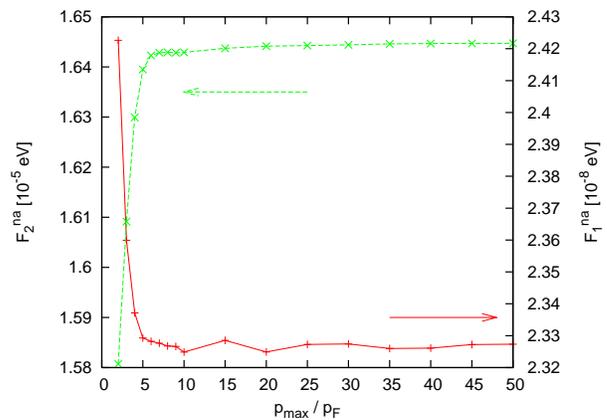,angle=-90,width=8cm}
  \caption{
    \label{fig_set_27}
    Convergence of $F_{1}^{na}$ and $F_{2}^{na}$ for Na at $T$ = 200 K.
  }
\end{figure}

\begin{widetext}
\begin{eqnarray}
  \label{eq_F_1_na_einstein}
  \frac{F_{1}^{na}}{N} & = &
  \frac{ V_{A}^{2} }{ \left( 2 \pi \right)^{6} }
  \,\,
  \mathcal{P}
  \int d\vec{p}
  \int d\vec{p'}
  \,\,
  \frac{\hbar^{2}}{M}
  \hbar \omega_{E}
  \left( n_{E} + \frac{1}{2} \right)
  \frac{f_{\vec{p}}}{\epsilon_{\vec{p}} - \epsilon_{\vec{p'}}}
  \frac{\left| \vec{p'} - \vec{p} \right|^{2}
    \left[ U \left( \vec{p'} - \vec{p} \right) \right]^{2}}
    {\left[ \epsilon_{\vec{p}} - \epsilon_{\vec{p'}} \right]^{2}
     - \left[ \hbar \omega_{E} \right]^{2}} \\
  \label{eq_F_2_na_einstein}
  \frac{F_{2}^{na}}{N} & = &
  \frac{ V_{A}^{2} }{ \left( 2 \pi \right)^{6} }
  \,\,
  \mathcal{P}
  \int d\vec{p}
  \int d\vec{p'}
  \,\,
  \frac{\hbar^{2}}{2 M}
  f_{\vec{p}} \left( 1 - f_{\vec{p'}} \right)
  \frac{\left| \vec{p'} - \vec{p} \right|^{2}
    \left[ U \left( \vec{p} - \vec{p'} \right) \right]^{2}}
    {\left[ \epsilon_{\vec{p}} - \epsilon_{\vec{p'}} \right]^{2}
     - \left[ \hbar \omega_{E} \right]^{2}},
\end{eqnarray}
\end{widetext}

\noindent
where $\mathcal{P}$ stands for the principal part. We would like to mention that
neither one of the two non--adiabatic contributions has any obvious upper limits
for the integration on $\vec{p'}$ and we need to make sure that we understand
the convergence behavior of their integrands for large $\left| \vec{p'}
\right|$. The pseudopotential, $U (\vec{q})$, converges as $q^{-2}$ for large
$q$ and we conclude that the second non--adiabatic contribution, $F_{2}^{na}$,
converges as $(p')^{-4}$. Due to the additional factor of $1/\left( \epsilon_{p} -
\epsilon_{p'} \right)$ in $F_{1}^{na}$, the first non--adiabatic contribution
converges as $(p')^{-6}$.  We can therefore expect good convergence behavior
for the two non--adiabatic contributions in terms of the integration over
$\vec{p'} = \vec{p} + \vec{k} + \vec{Q}$. Fig.  \ref{fig_set_27} shows this
nicely. $p_{\mathrm{max}}$ is the upper integration limit of $p'$.

The 6--dimensional integration of eqs. (\ref{eq_F_1_na_einstein}) and
(\ref{eq_F_2_na_einstein}) was done using the VEGAS method
\cite{Lepage_80:VEGAS, Press_NRC_88}. We took special care to ensure convergence
around the divergent poles of the integrands when calculating the principal
part. This was done by first reducing the poles from second to first order and
then removing the divergence by subtracting a suitably chosen function from the
integrand and integrating its pole by hand. This effectively smoothed the
integrand sufficiently so that our VEGAS calculation converged much more
rapidly. It also improved the reliability of the VEGAS error estimate and
therefore the accuracy of our calculation. Since we are using spherical
coordinates, this procedure has the advantage that we can very efficiently limit
the integration region around the Fermi surface by using appropriate integration
limits. It is also simpler to implement than a tetrahedron method in which the
Brillouin zone is divided into tetrahedra which are then summed up.

\begin{figure}
  \psfig{file=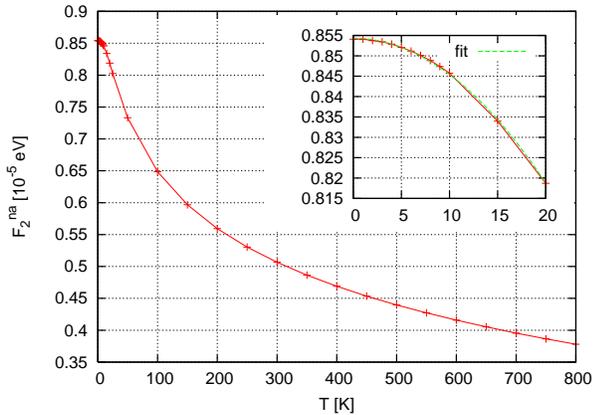,angle=-90,width=8cm}
  \caption{
    \label{fig_set_09}
    $F_{2}^{na}$ for K, Harrison model.
  }
\end{figure}

\subsection{
  Non--Adiabatic Contribution to $F_{ep}$ for different materials
}

We calculated $F_{1}^{na}$ for Na and $F_{2}^{na}$ for Na, K, Al, and Pb. Table
\ref{table_material_constants} shows the material specific constants used in our
calculations.

\begin{table}
\caption{
  \label{table_material_constants}
  Material constants
}
\begin{tabular}{lcccc}
\hline
\hline
  & 
    Na &
    K &
    Al &
    Pb \\
\hline
  structure &
    bcc &
    bcc &
    fcc &
    fcc \\
  $T_{\rho}$ [K] &
    90 &
    9 &
    80 &
    80 \\
  $\rho$ [g cm${}^{-3}$] &
    1.005 &
    0.904 &
    2.731 &
    11.55 \\
  $\left< \hbar \omega \right>$ [meV] &
    10.53 &
    6.42 &
    25.79 &
    5.90 \\
  a [\AA] &
    4.23 &
    5.24 &
    4.05 &
    4.95 \\
  $p_{F}$ [\AA${}^{-1}$] &
    0.92 &
    0.74 &
    1.75 &
    1.57 \\
  $\epsilon_{F}$ [eV] &
    3.24 &
    2.11 &
    11.66 &
    9.45 \\
  $T_{melt}$ [K] &
    407 &
    368 &
    1234 &
    722 \\
\hline
\hline
\end{tabular}
\end{table}

\begin{table}
\caption{
  \label{table_pseudopotential_parameters}
  The pseudopotential parameters. Parameters for the Harrison model are
  $\hat{\rho}$ and $\hat{\beta}$, while the single parameter for the Ashcroft
  model is $\bar{\rho}$.
}
\begin{tabular}{lcccc}
\hline
\hline
  &
    Na &
    K &
    Al &
    Pb \\
\hline
  Reference &
    \citep[p. 406]{Wallace_72:TOC} &
    \citep[p. 406]{Wallace_72:TOC} &
    \citep[p. 415]{Wallace_72:TOC} &
    \citep[p. 502]{Parks_Superconductivity_69} \\
  $Z$ &
    1 &
    1 &
    3 &
    4 \\
  $\xi$ &
    1.81 &
    1.77 &
    1.90 &
    1.81 \\
  $\hat{\rho}$ [$a_{B}$] &
    0.50 &
    0.69 &
    0.24 &
    --- \\
  $\hat{\beta}$ [Ry $a_{B}^{3}$] &
    37 &
    66 &
    47.5 &
    --- \\
  $\bar{\rho}$ [$a_{B}$] &
    --- &
    --- &
    1.117 &
    1.12 \\
\hline
\hline
\end{tabular}
\end{table}

Figs. \ref{fig_set_06} and \ref{fig_set_05} show our results for Na using a
single phonon mode at $\hbar \omega_{E} = 10.53$ meV for $F_{1,2}^{na}$.
$F_{1}^{na}$ exhibits linear temperature dependence for higher temperatures and
even at temperatures as low as 50 K we hardly see deviations from linear
behavior. This confirms our previous estimate of this contribution's temperature
dependence based on a phase space argument. In section
\ref{sec_analytic_temp_dependence} we had argued that when we rewrite the
Fermi--Dirac distribution factor in eq. (\ref{eq_F_1_na_einstein}) as eq.
(\ref{eq_f}) we expect the $g_{\vec{p}}$--factor to be dominant due to its much
larger phase space. Since $g_{\vec{p}}$ is not temperature dependent, the only
temperature dependent factor is the phonon distribution which is linear at high
temperatures. As we had argued in the same section also, the magnitude of
$F_{1}^{na}$ is much smaller than the magnitude of $F_{2}^{na}$.  Based on this
magnitude difference we will neglect $F_{1}^{na}$ from this point onward.

Fig. \ref{fig_set_05} shows $F_{2}^{na}$ and as expected this contribution is
positive all throughout the temperature regime shown. Starting from $T = 0$ the
curve has negative curvature but quickly switches to positive curvature and
never crosses zero even at high temperatures. From the inset in Fig.
\ref{fig_set_05} we get the values of $C_{2}$ and $A_{2}$ listed in Table
\ref{table_results}.

\begin{figure}
  \psfig{file=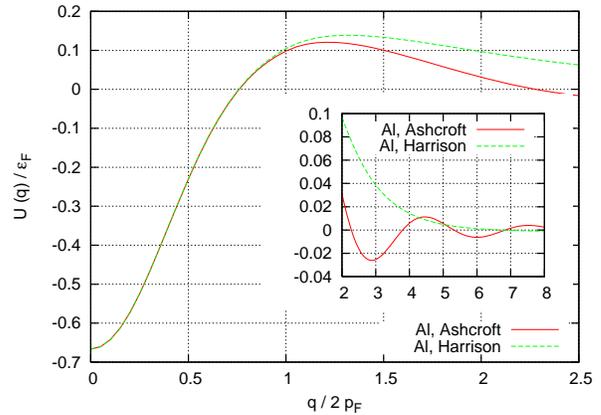,angle=-90,width=8cm}
  \caption{
    \label{fig_pseudopotential_Al}
    Harrison and Ashcroft pseudopotential models for Al. The parameters are
    given in Table \ref{table_pseudopotential_parameters}.
  }
\end{figure}

Fig. \ref{fig_set_09} shows our result for $F_{2}^{na}$ for K. The magnitude of
the $F_{2}^{na}$ contribution is smaller by about a factor of 3 compared to Na.
The difference is presumably largely due to the difference in $\epsilon_{F}$
which affects the size of the pseudopotential $U (q)$. Since the ratio of Fermi
energies is about 3/2 and the pseudopotential enters the free energy squared, we
get about a factor of 2 difference between the two materials just based on this
term.

\begin{figure}
  \psfig{file=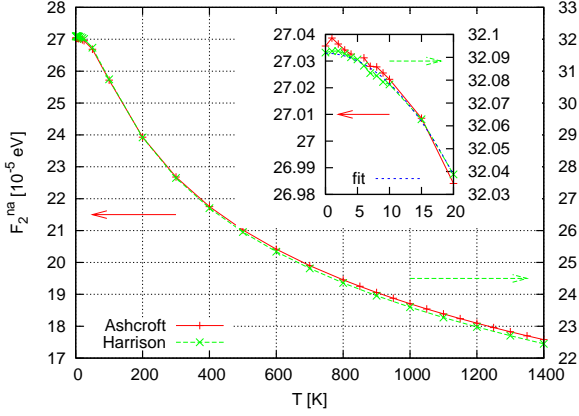,angle=-90,width=8cm}
  \caption{
    \label{fig_set_21}
    $F_{2}^{na}$ for Al for different pseudopotential models.
  }
\end{figure}

The pseudopotentials for Al are shown in Fig. \ref{fig_pseudopotential_Al}.
Between zero and $q / (2 p_{F}) \approx 1$ the two models are identical. Between
$1 \lesssim q / (2 p_{F}) \lesssim 4$ the Harrison model is larger than the
Ashcroft model and for $q / (2 p_{F}) \gtrsim 4$ both models slowly approach
zero.  Fig. \ref{fig_set_21} shows our results for Al which we calculated for
these two pseudopotential models. $F_{2}^{na}$ using an Ashcroft pseudopotential
model is shifted by about 0.05 meV to lower energies compared to the Harrison
model. The qualitative temperature dependence is not affected.  The difference
between the two models in terms of their convergence to zero for large $q$
therefore does not appear to be relevant in affecting the temperature dependence
of $F_{2}^{na}$. The difference in magnitude for intermediate $q$ does enter the
overall magnitude of $F_{2}^{na}$ and makes a difference of about 20 - 30\%. The
inset Fig. \ref{fig_set_21} shows a fit using eq.  \ref{eq_2_na_T_dependence},
and the results for $C_{2}$ and $A_{2}$ are listed in Table \ref{table_results}.

\begin{figure}
  \psfig{file=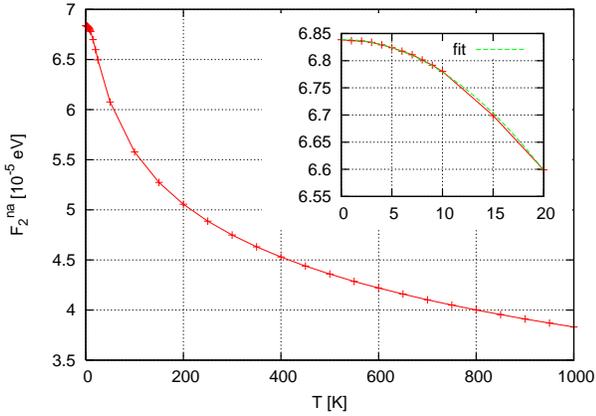,angle=-90,width=8cm}
  \caption{
    \label{fig_set_13}
    $F_{2}^{na}$ for Pb, Ashcroft pseudopotential model.
  }
\end{figure}

\begin{figure}
  \psfig{file=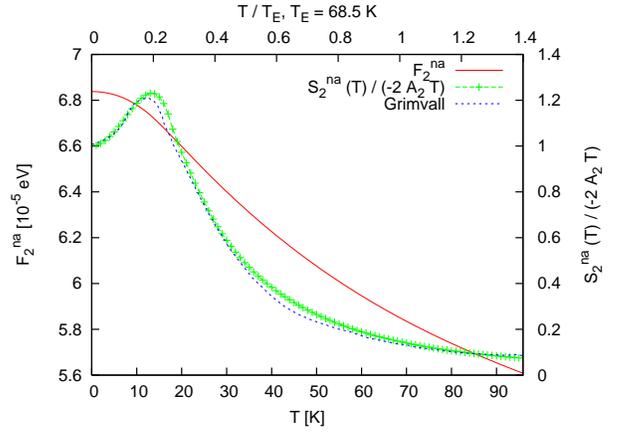,angle=-90,width=8cm}
  \caption{
    \label{fig_set_29}
    $F_{2}^{na}$ and $S_{2}^{na} (T) / \left( -2 \, A_{2} T \right)$, where
    $A_{2}$ was defined in eq. (\ref{eq_2_na_T_dependence}), compared with
    earlier results by Grimvall on Pb. $F_{2}^{na}$ is plotted with axes on the
    left and bottom, $S_{2}^{na}$ and Grimvall's results are plotted with axes
    on the right and top.
  }
\end{figure}

Fig. \ref{fig_set_13} shows our results for Pb. In Fig. \ref{fig_set_29} we show
the entropy calculated from the free energy and compare our result with earlier
results by \citet[Fig. 5.19]{Grimvall_EPI_81}. Our results are scaled with an
Einstein temperature of $\Braket{ \hbar \omega } = 5.9 \mbox { meV} = 68.5 \mbox
{ K}$.  We find very good agreement with Grimvall's results. Grimvall calculated
the total entropy, i.e. $S_{Grimvall} = S^{ad} + S_{1}^{na} + S_{2}^{na}$ for Pb
up to $T = 1.4 \,\, T_{E}$.  Up to this temperature, our purely non--adiabatic
entropy, $S_{2}^{na}$, agrees very nicely with his results \footnote{To remind
the reader we would like to note that as pointed out previously, the first
non--adiabatic contribution, $F_{1}^{na}$, and consequently its related entropy,
$S_{1}^{na}$, is negligible compared to the other two contributions.}.
The adiabatic contribution remains small compared to the non--adiabatic up to
much higher temperatures. We shall address this in a future publication.

\begin{table}
\caption{
  \label{table_results}
  $F_{2}^{na} (T = 0)$ and curvatures at low--$T$ (Units: $C_{2}$ [$10^{-8}$
  eV/atom], $A_{2}$ [$10^{-5}$ eV/atom K${}^{2}$])
}
\begin{tabular}{llcccc}
\hline
\hline
  &
  &
    Na &
    K &
    Al &
    Pb \\
\hline
  \multirow{2}{*}{$C_{2}$} &
    (Harrison) &
    2.270 &
    0.8543 &
    32.09 &
    --- \\
  &
    (Ashcroft) &
    --- &
    --- &
    27.04 &
    6.839 \\
\hline
  \multirow{5}{*}{$A_{2}$} &
    (expt. \footnote{$A_{2}$ is calculated from experiment using
    low--temperature specific heat data. From eq. (\ref{eq_2_na_T_dependence})
    we find $C_{ep} = - 2 A_{2} T$. At low $T$, $C_{expt} = \Gamma_{expt} T$ and
    $C_{th} = \Gamma_{bs} \left( 1 + \lambda \right) T$. Our $C_{ep} = \lambda
    \, \Gamma_{bs} \, T$ and we find $A_{2} = -\frac{1}{2} \left( \Gamma_{expt}
    - \Gamma_{bs} \right)$. $\Gamma_{expt}$ taken from \citet[Table on p.
    157]{Kittel_96:ISSP}. The analysis was done by \citet[Table I and eq.
    (6)]{Allen_87}.}) &
    -0.10 &
    -0.13 &
    -0.19 &
    -0.86 \\
  &
    \citep[Table 27]{Wallace_72:TOC} &
    -0.14 &
    -0.14 &
    -0.30 &
    --- \\
  &
    (ppt \footnote{$A_{2}$ is calculated from pseudopotential perturbation
    theory, and given by $A_{2} = -\frac{1}{2} \Gamma_{bs} \lambda$. $\lambda$
    is taken from the data collection of \citet[Tables III, IV, and
    V]{Grimvall_76}.}) &
    -0.10 &
    -0.12 &
    -0.24 &
    -0.91 \\
  &
    (Harrison) &
    -0.0878 &
    -0.0889 &
    -0.1316 &
    --- \\
  &
    (Ashcroft) &
    --- &
    --- &
    -0.1285 &
    -0.6039 \\
\hline
\hline
\end{tabular}
\end{table}

\section{
  Conclusions
}

A thorough analysis of the first non--adiabatic term, $F_{1}^{na}$, reveals a
ground state contribution, $\propto g_{\vec{p}}$, and a contribution due to
excited electronic states, $\propto \left( f_{\vec{p}} - g_{\vec{p}} \right)$,
cf. eq. (\ref{eq_f}). We find that the ground state term, since it's temperature
dependence is determined by the phonon factor alone, leads to a $T$--dependence
of $F_{1}^{na}$ as given in eq. (\ref{eq_T_dependence}). Electronic excitations
above the ground state, the $\left( f_{\vec{p}} - g_{\vec{p}} \right)$ term,
will lead to a temperature dependence which is determined by the leading terms
in a Sommerfeld expansion and give a $T^{2}$--dependence at low temperatures for
$F_{1}^{na}$. Using a phase space argument we found that the ground state term
will dominate the temperature dependence and essentially determine it. The
contribution from the excited electronic states to the temperature dependence
will be much smaller than the ground state contribution and not visible in our
calculations. We also expect the magnitude of $F_{1}^{na}$ to be $\propto \left(
\Braket{\hbar \omega} / \epsilon_{F} \right) F_{2}^{na}$ and hence negligible
compared to $F_{2}^{na}$.  Our numerical results confirm our suspicions, cf.
Fig.  \ref{fig_set_06}. The calculated temperature dependence is linear, i.e.
the ground state term dominates the temperature dependence of $F_{1}^{na}$. The
absolute magnitude of $F_{1}^{na}$ is by about three orders of magnitude smaller
than what we find for $F_{2}^{na}$ and we find that we can neglect this
contribution.

The more complicated dependence of the Fermi--Dirac factor does not allow us to
split $f_{\vec{p}} \left( 1 - f_{\vec{p'}} \right)$ into parts as in eq.
(\ref{eq_f}), and the low--temperature dependence of $F_{2}^{na}$ is determined
by the leading terms in a Sommerfeld expansion. The $T$--dependence therefore is
given by eq. (\ref{eq_2_na_T_dependence}). Studying the sign of the single
factors in $F_{2}^{na}$ we find that this contribution is positive for all
temperatures. Starting from a non--zero constant at $T = 0$ the magnitude slowly
approaches zero as the temperature increases. Our numerical results confirm this
analysis, cf. Fig. \ref{fig_set_05}. The constant $C_{2}$ is the non--adiabatic
correction to the adiabatic ground state energy \cite{Wallace_02:SPCL},
$\Phi_{0} (V)$. In general $C_{2} \ll \Phi_{0}$ and is negligible. The curvature
$A_{2}$ represents the leading electron--phonon correction to the
low--temperature free energy and specific heat. The latter can be written as

\begin{equation}
  C = \frac{\pi^{2}}{3} N_{bs}
    k_{B}^{2}
    \left( 1 + \lambda \right) T,
\end{equation}

\noindent
where the electronic density of states given by bandstructure alone, $N_{bs}$,
comes from $\mathcal{H}_{el}$ and is corrected by the electron--phonon
interaction with the factor $\lambda$. This factor is measurable in experiments
and can get quite large for certain metals. In the case of lead for instance,
$\lambda \approx 1.2$. We calculated the constant $C_{2}$ and the curvature
$A_{2}$ for all four metals we studied and our results are listed in Table
\ref{table_results}. In addition to our results we also included results from
experiment and from previous theoretical studies. Our results agree fairly well
with the other numbers. It should be pointed out that while $F_{2}^{na}$ is
important for low temperatures, it is negligible at high--$T$. Comparing the
entropy from the non--adiabatic contribution with the electronic contribution
for instance, the ratio $S_{2}^{na} / S_{el}$ while important at low
temperatures, becomes negligible for $T / T_{E} \gtrsim 1$.

Our results in Table \ref{table_results} indicate variations in the different
theoretical calculations. But regardless of the details of the theoretical
description, they show the ability of the theory to calculate the experimental
results. Since we are working within the Einstein approximation, they also
provide confirmation of our previous assumption that the Einstein model will
give us the correct temperature dependence and accurate results.

\appendix

\section{
  \label{appendix_calculation_details}
  Technical Details regarding the calculation
}

We used a diagrammatic technique to calculate the electron--phonon contribution
eq. (\ref{eq_F_ep}). To leading order in the interaction however this approach
is completely equivalent to a standard perturbation theory approach taken by
Wallace for instance \citep[Section 25]{Wallace_72:TOC}. We used a linked
cluster expansion, given by

\begin{equation}
  \label{eq_starting_Hamiltonian}
  \Omega - \Omega_{0} =
    \frac{1}{2 \beta} \int_{0}^{1} \frac{ d\eta }{ \eta }
    \sum_{\vec{k} \sigma} \sum_{i k_{n}} \Sigma^{\eta} (\vec{k} \sigma, i k_{n})
    \mathcal{G}^{\eta} (\vec{k} \sigma, i k_{n}),
\end{equation}

\noindent
where the self--energy and Green's function implicitly depend on the coupling
constant \cite{Fetter_71:Quantum_Theory_of_Many--Particle_Systems}, $\eta$.
Since we are using a plane wave basis and a free electron dispersion, we need to
include the zeroth order term in $\mathcal{H}_{ep}$. The electron self energy,
$\Sigma (\vec{k}, i k_{n})$, includes diagrams of the form

\begin{equation}
  \includegraphics*[width=8cm]{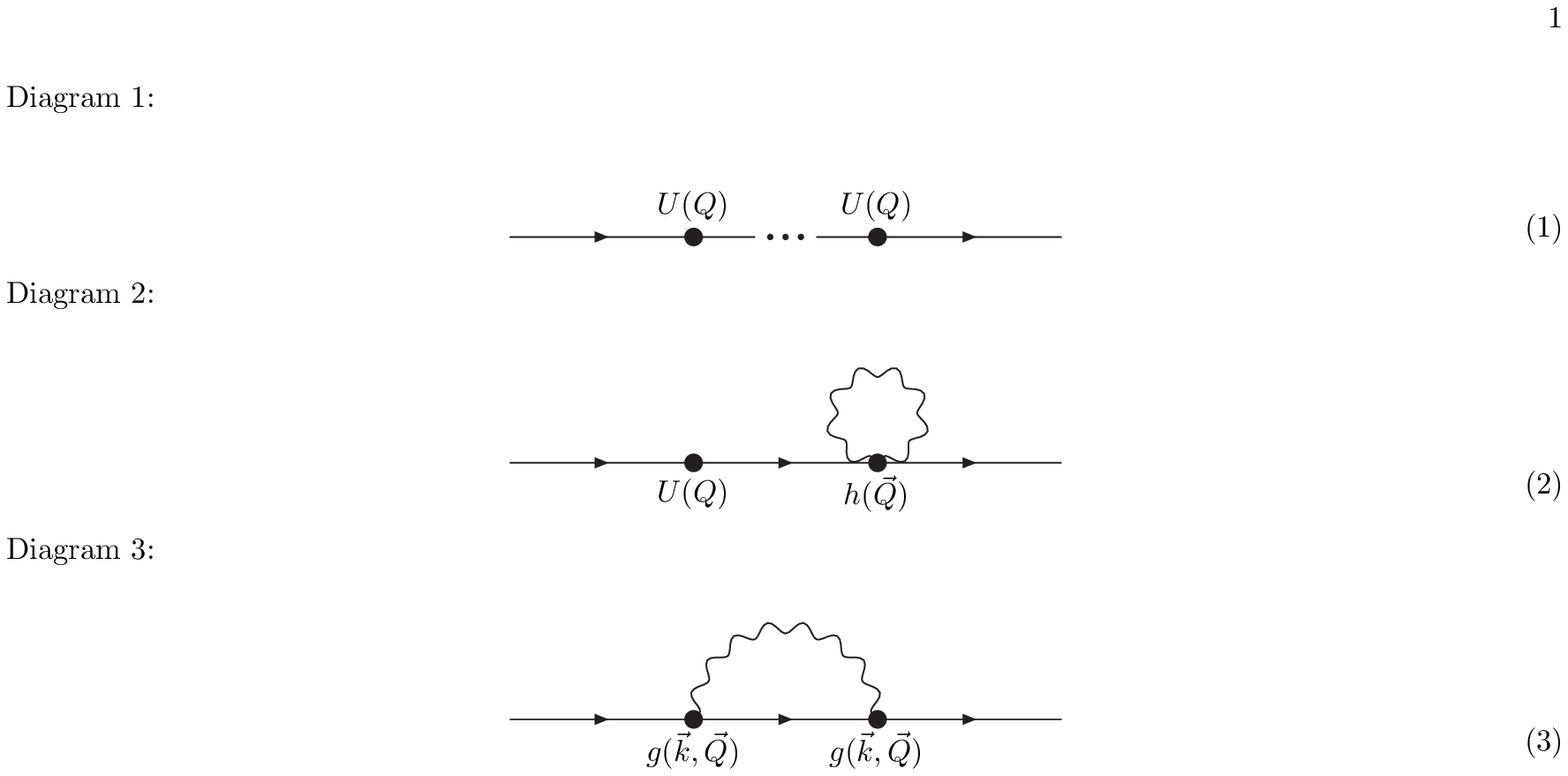}
\end{equation}

\noindent
in n${}^{th}$ order. We therefore do not consider these terms in
$\mathcal{H}_{ep}$. We do need to correct our plane wave basis and include terms
of the form

\begin{equation}
  \includegraphics*[width=8cm]{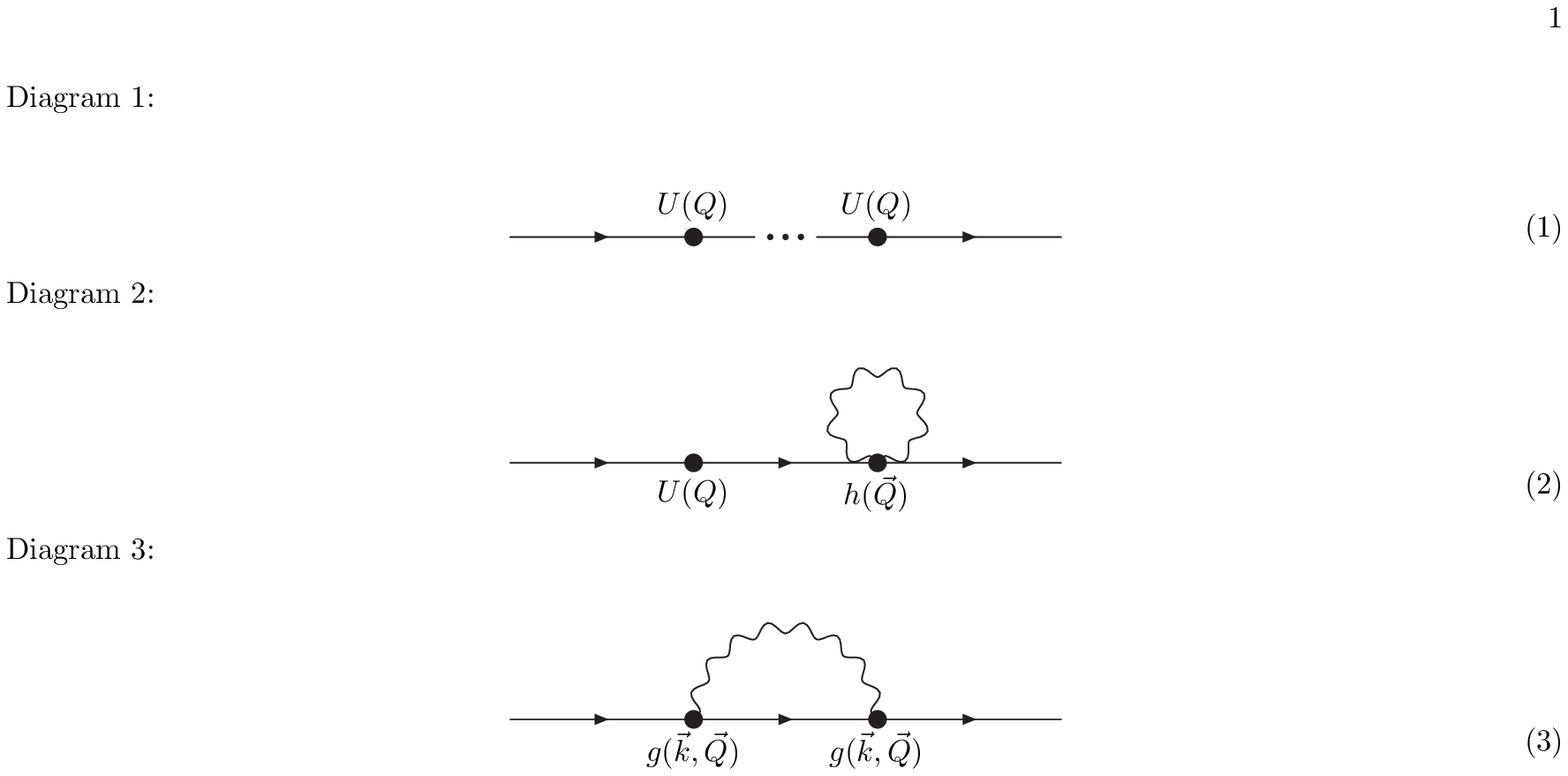}
\end{equation}

\noindent
This can be achieved in regular perturbation theory by correcting the wave
function to first order and then calculating the diagonal matrix elements in
first order. Eq. (25.17) of \citet{Wallace_72:TOC} corresponds to this kind of
diagram. The first term in (25.21) therefore is given by

\begin{equation}
  \includegraphics*[width=8cm]{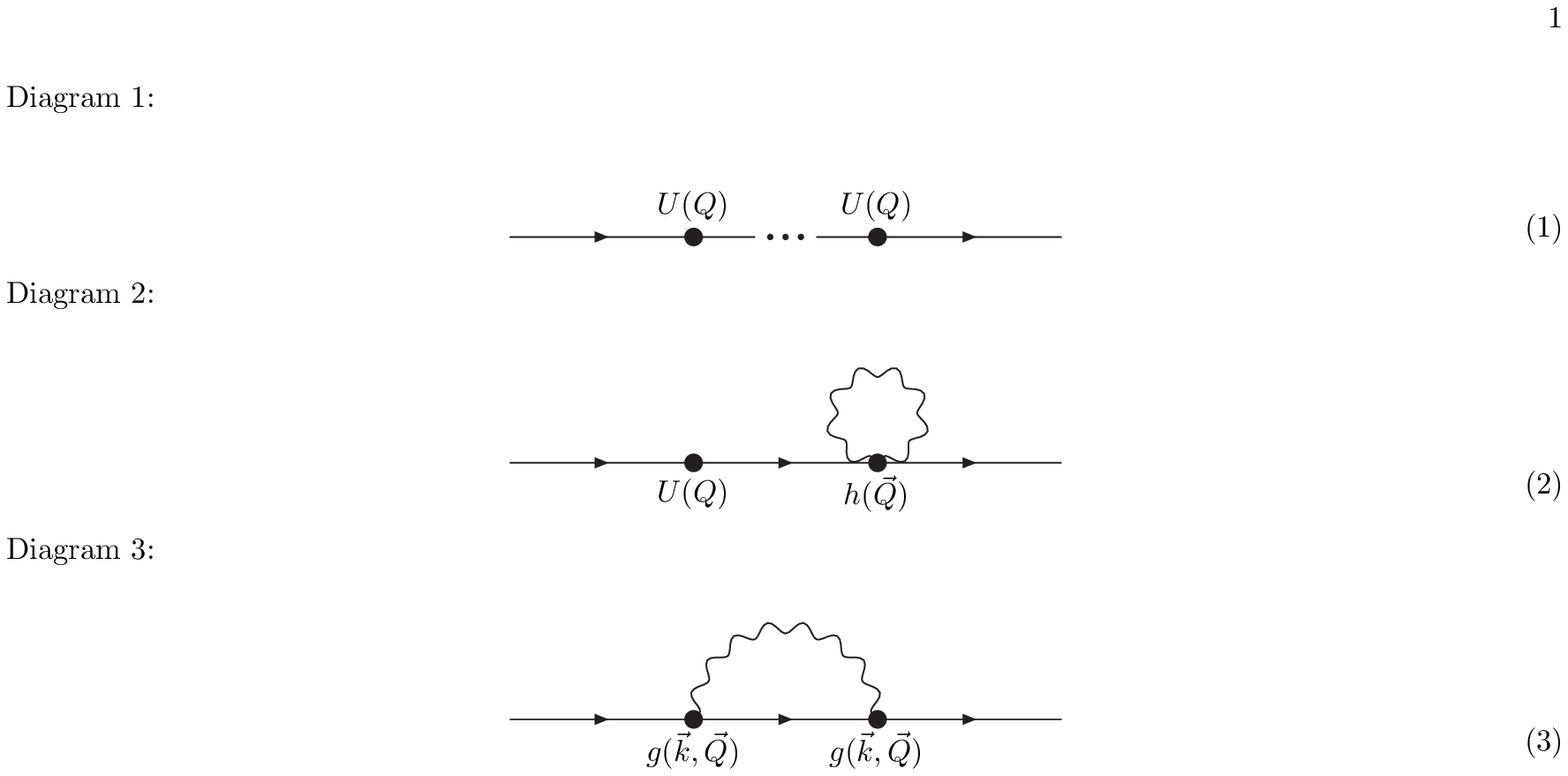}
\end{equation}

\bibliography{Paper}
\bibliographystyle{apsrev}

\end{document}